# THE PECULIARITIES OF EMISSION OF He(Z=2) PROJECTILE FRAGMENTS IN THE INTERACTIONS OF GOLD NUCLEI WITH Ag, Br EMULSSION NUCLEI AT THE ENERGIES OF 10.6 $A$GeV

**U. U. Abdurakhmanov, V. S. Navotny**

*Physical-Technical Institute, Scientific Association "Physics-Sun", Academy of Sciences of Uzbekistan, 2-Bodomzor yoli street, Tashkent, Uzbekistan, 100084*

**Abstract**

*In the interactions of gold nuclei with the energy 10.6 AGeV with the Ag, Br emulssion nuclei at the intermadiate impact parametes the dependencies of characteristics of individual emission acts of He-fragmetns on the centrality of the interaction and the degree of nucleus disintegration.*

**Key words:** avarage transverse momentum, He-fragment, projectile, target, interactions.

## 1. Experiment

In a nuclear emulsion irradiated at the BNL / AGS accelerator with gold nuclei with a kinetic energy of 10.6 $A$GeV, 1057 inelastic interactions were found. In each event, depending on ionization, all tracks of secondary charged particles emitted during the interaction were divided into the following categories: black (*b*-tracks), gray (*g*-tracks), relativistic (*s*-tracks), and fragments of the projectile with charge $z \geq 2$. Criteria for the separation of different categories of tracks are given, for example, in [1]. For all listed categories of particles, polar $\theta$ and azimuthal $\varphi$ emission angles were measured, and charges of multiply charged ($z \geq 2$) fragments of the projectile were determined. The fragments of the projectile (group *f*) included fragments of a projectile with a charge and singly charged fragments of the projectile, which are *s*-particles with angles of emission $\theta < 1°$. Fragments of the projectile with a charge $z = 2$ were identified visually. The ionization on the tracks of such particles is constant over a large length and is equal to $I \approx 4I_0$, where $I_0$ is the minimum ionization on tracks of singly charged particles. To determine the charges of the fragments with $z \geq 3$, we used measurements of the density of electrons at a length of at least *10* mm; Calibration was carried out on the primary traces of a known charge and fragments with $z = 2$. The accuracy of the determination of the charges for the fragments with $z < 10$, $10 \leq z < 28$, $28 \leq z < 40$ and $z \geq 40$ was $\pm 0.5$, $\pm 1$, $\pm 2$, $\pm 3$ respectively.

For the analysis, 340 events of inelastic interaction of gold nuclei with Ag and Br emulsion nuclei were selected at the for medium impact parameters. These events met the following selection criteria:

1. the number of fragments of the target $n_g + n_b \geq 8$;
2. The number of fragments of the shell with a charge $z = 2$   $N_{(z=2)} \geq 3$.

In the selected events, the number of fragments of the projectile varied from $3$ up to $16$. The average multiplicity of $He_{(z=2)}$-fragments of the projectile in the ensemble of events under consideration was $\langle n_{He} \rangle = 5.75 \pm 0.12$.

## 2. The average transverse momentum per nucleon of the He projectile fragments in an individual event

The formation of the He-fragments of the projectile was studied by means of quantities characterizing individual events. One of these quantities is the average transverse momentum per nucleon of the H- fragments of the projectile in an individual event

$$\langle (P_T / A)_{He} \rangle = \frac{\sum_{i=1}^{n}(P_{T_i} / A)_{He}}{n}, \qquad (1)$$

where $n$ is the number of He fragments in the individual event. Assuming that each He-fragment of the projectile has a longitudinal momentum per nucleon as the primary nucleus, the magnitude of the transverse momentum per nucleon of the $i$-th He-fragment of the projectile is defined as $(P_{T_i}/A)_{He} = P_0 \sin\theta_i$ where $\theta_i$ is the polar emission angle of He-fragment and $P_0$ is the momentum per nucleon of the incident nucleus.

The experimental distribution of the mean transverse momentum per nucleon $\langle(P_T/A)_{He}\rangle$ of the He-fragments of the projectile in an individual event, shown in Fig. 1, does not agree with the distribution obtained within the FRITIOF-M model [2].

In the model FRITIOF-M, a complete simulation of the interactions of gold nuclei with an energy of 10.6 $A$GeV with the emulsion nuclei and the generated events were subjected to selection simulating the experimental situation. The FRITIOF-M model is a combined model where the first stage of the reactions is played out using the phenomenological approach [3] used in the modified version of the FRITIOF model [4], and the decay of the residue nuclei is according to the statistical model of multifragmentation of nuclei [5]. The FRITIOF-M model takes into account the energy-momentum conservation laws and assumes a statistically isotropic decay of residual nuclei, but there are no correlations due to some other reasons.
Table 1 shows the values

$$\langle\langle(P_T/A)_{He}\rangle\rangle = \frac{1}{N}\sum_{i=1}^{N}\langle(P_T/A)_{He}\rangle_i \qquad (2)$$

averaged over $N$ events of the ensemble of values of the mean transverse momenta per nucleon of the He-fragments of the projectile in individual acts of interaction of gold nuclei with an energy of 10.6 $A$GeV with emulsion nuclei at medium impact parameters and a number of subensembles giving information on the dependence $\langle\langle(P_T/A)_{He}\rangle\rangle$ on the impact parameter of interaction (characterized by the magnitude of the bound charge residual nucleus $z_{bound}$, i.e., the total charge of the projectile fragments with charge $z \geq 2$) and the degree of disintegration of the fragmenting core (characterized by the maximum charge $z_{max}$ of the products of its disintegration). As it can be seen from Table 1, the experimental value $\langle\langle(P_T/A)_{He}\rangle\rangle$ increases with the transition to more central interactions (with $z_{bound}$ decreasing), and also increases with $z_{max}$ decreasing, while in the FRITIOF-M model this value decreases with decreasing impact parameter.

One of the reasons for the disagreement between the shape of the model and experimental distributions (Figure 1) and of the experimental value $\langle\langle(P_T/A)_{He}\rangle\rangle$ which is higher than the model value (see Table 1) is apparently the collective motion of the residual nucleus as a whole in the transverse interaction plane not taken into account in the model FRITIOF-M.

### 3. The asymmetry index of transverse momentum values of He fragments in an individual event

In addition to the mean transverse momentum, the magnitude $g_1'$ of the asymmetry of the arrangement of He-fragments in the scale of the values of their transverse momenta can serve as another quantity characterizing the individual act of emission of He-fragments [6]:

$$g_1' = \frac{\sqrt{n-1}}{n-2}\frac{m_3}{\sqrt{m_2^3}}, \qquad (3)$$

where $m_l$ is the central momenta of $l$-th order.

$$m_l = \frac{1}{n}\sum_{i=1}^{n}(x_i - \langle x\rangle)^l, \; \langle x\rangle = \frac{1}{n}\sum_{i=1}^{n}x_i, \qquad (4)$$

$n$ is the number of He-fragments in an event; $x_i = (P_{T_i}/A)_{He} = P_0 \sin\theta_i$ is the magnitude of the transverse momentum per nucleon of the $i$-th He-fragment of the projectile.

If in an event $(n-1)$ He-projectile fragments have the same values of the quantity $(P_{T_i}/A)$ equal to $a$, and for one He-fragment $(P_{T_i}/A) = b$, then

$$g'_1 = 1 \quad \text{for} \quad a < b$$
$$g'_1 = -1 \quad \text{for} \quad a > b \qquad (5)$$

for any values of $a$ and $b$ ($a \neq b$) independently of $n$.

The above mentioned values (5) of the quantity $g'_1$ correspond to events with the highest degree of asymmetry of the values of transverse momenta. With a symmetrical arrangement of the values of transverse momenta, the values $g'_1 = 0$. In Fig. 2 shows the characteristic patterns of the arrangement of He-fragments of projectiles in the scale of transverse momenta of an individual event in the case of extreme asymmetry (a, b) and in cases of symmetry (c, d, e).

Note that the quantity $g'_1$ is parametrically invariant, i.e. it does not depend on the shift of the origin and on the scale of the transverse moment scale of the individual event. The experimental distribution of events by the quantity $g'_1$ is confined in the interval of $-1 \div +1$.

In the distribution of experimental events by the magnitude of $g'_1$, the events with positive values of $g'_1$ prevail (Fig.3), i.e. events with such asymmetry, when the majority of He fragments are shifted to the beginning of the scale ordered in the order of increasing of their transverse momenta of the individual event. As it can be seen from given in Table 1 values of $\langle g'_1 \rangle$ averaged over $N$ events of the ensembles, the asymmetry effect relative to $g'_1 = 0$ of the experimental distribution by $g'_1$ is enhanced with increasing of the centrality of the collision. The shape of the distribution by the magnitude $g'_1$ in the model FRITIOF-M (Fig. 3) does not correspond to the experimental one.

## 4. Conclusion

In the interactions of gold nuclei with an energy of 10.6 *A*GeV with Ag, Br emulsion nuclei with average impact parameters, the dependences of the characteristics of the individual acts of emission of the He fragments of the projectile on the centrality of the interaction and the degree of disintegration of the nucleus were studied. It is shown that the average value of the transverse momentum of the He-fragments of the projectile per nucleon increases with the transition to more central events, and also increases with the decrease in the degree of disintegration of the nucleus, while in the FRITIOF model this value decreases with the decrease of the impact parameter.

The authors are grateful to the participants of EMU-01 collaboration for joint work on collecting and analyzing the experimental data.

## References

[1] M. I. Adamovich *et al.* (EMU-01 Collab.), Eur. Phys. J. A **2**, 61 (1998).
[2] M. I. Adamovich *et al.* (EMU-01/12 Collab.), Z. Phys. A **359**, 277 (1997).
[3] Kh. El-Waged, V. V. Uzhinskii, ЯФ **60**, 925 (1997).

[4] M. I. Adamovich *et al.* (EMU-01/12 Collab.), Z. Phys. A **358**, 337 (1997).
[5] J. P. Bondorf, A. S. Botvina *et al.*, Phys. Rep. **257**, 133 (1995).
[6] Ш. Абдужамилов и др., ЯФ **25**, 575 (1977).

**Table 1**: The values of the mean transverse momenta per nucleon of the He fragments of the projectile of the individual event averaged over $N$ events of the ensemble, and also $\langle (g'_1)_{He} \rangle$ - the mean values of the asymmetry indices of the values of transverse momenta

| Ensemble | $\langle\langle (P_T/A)_{He} \rangle\rangle$, MeV/c | | $\langle (g'_1)_{He} \rangle$ | |
|---|---|---|---|---|
| | experiment | FRITIOF-M | experiment | FRITIOF-M |
| весь | 116 ± 3 | 61 ± 1 | 0.21 ± 0.03 | 0.05 ± 0.04 |
| $6 \leq z_{bound} < 30$ | 131 ± 6 | 59 ± 1 | 0.30 ± 0.05 | 0.12 ± 0.04 |
| $30 \leq z_{bound} < 50$ | 109 ± 4 | 64 ± 1 | 0.18 ± 0.04 | -0.07 ± 0.07 |
| $z_{bound} \geq 50$ | 106 ± 4 | 70 ± 1 | 0.14 ± 0.05 | -0.28 ± 0.13 |
| $2 \leq z_{max} < 10$ | 126 ± 6 | 55 ± 1 | 0.28 ± 0.05 | 0.15 ± 0.06 |
| $10 \leq z_{max} < 30$ | 114 ± 4 | 64 ± 1 | 0.22 ± 0.04 | 0.07 ± 0.05 |
| $z_{max} \geq 30$ | 104 ± 4 | 66 ± 1 | 0.08 ± 0.06 | -0.24 ± 0.09 |

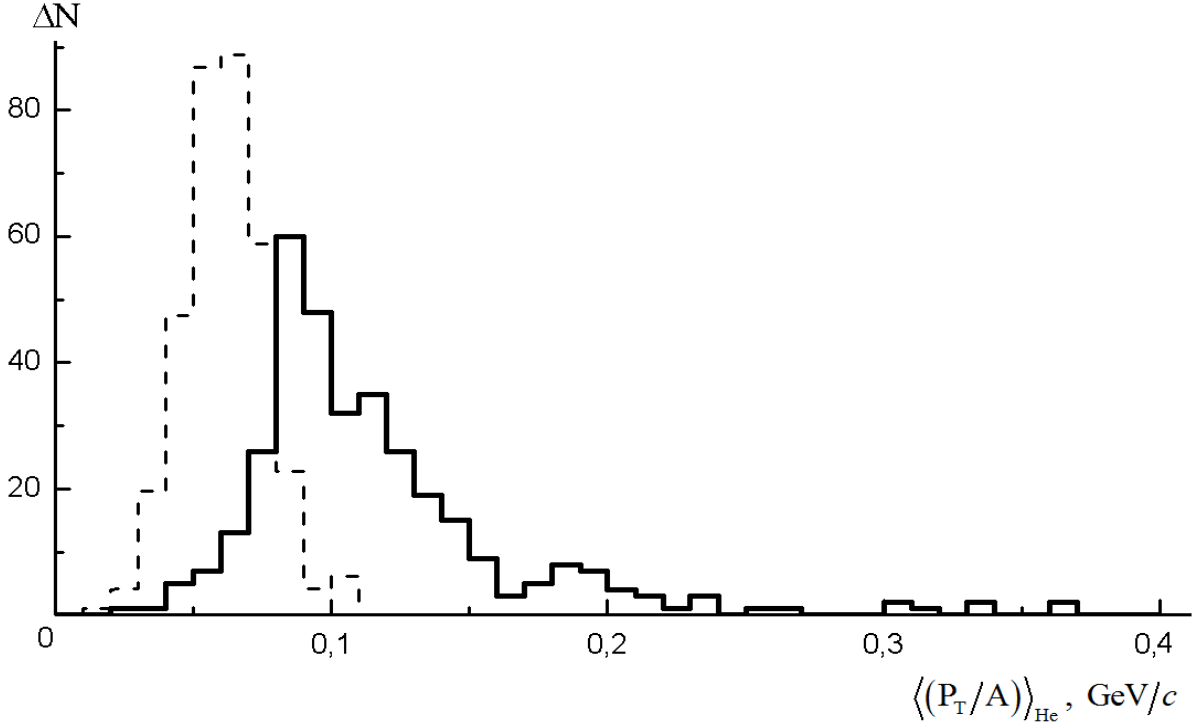

*Fig. 1. Distribution by the average transverse momentum per nucleon of the He-fragments of the projectile in individual acts of interaction of gold nuclei with an energy of 10.6 A GeV with Ag, Br emulsion nuclei with mean impact parameters. The solid line is an experiment, the dotted line is the FRITIOF-M model..*

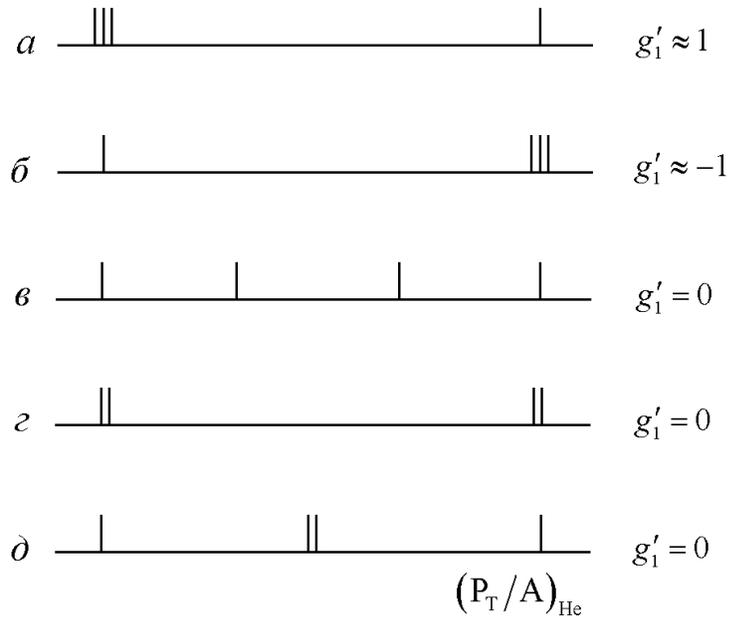

*Fig. 2. The examples of the arrangement of He fragments in the scale of the values of their transverse momenta of an individual event and the values of the asymmetry $g'_1$ corresponding to these configurations.*

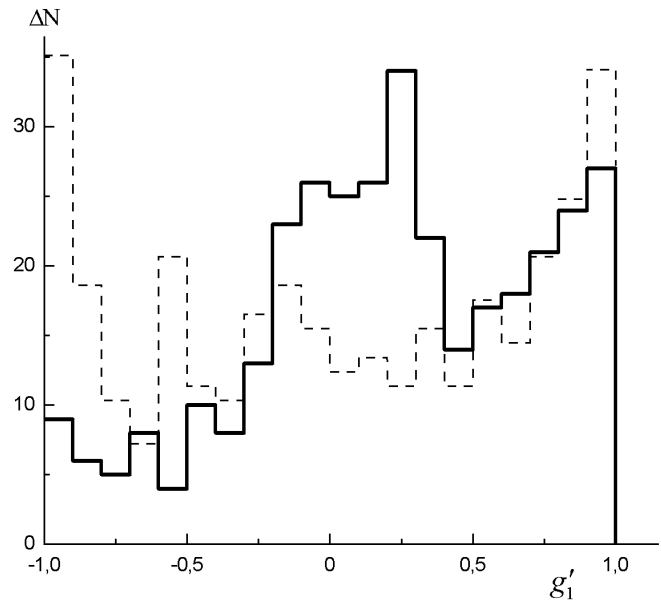

*Fig. 3. Distribution of events by the magnitude of $g'_1$ - the asymmetry index of the He-fragments of the projectile in the scale of the values of their transverse momenta. The solid line is an experiment, the dotted line is the FRITIOF-M model.*